\documentclass[prl,aps,twocolumn,superscriptaddress,floatfix,noshowpacs,nofootinbib,10pt]{revtex4-2}

\usepackage[utf8]{inputenc}     	
\usepackage{amsmath, amssymb, mathtools, mathrsfs}
\usepackage[colorlinks=true]{hyperref}
\hypersetup{colorlinks,linkcolor={red},citecolor={blue},urlcolor={blue}}  
\usepackage{color}
\usepackage{graphicx}
\usepackage[normalem]{ulem}
\usepackage{mathtools}
\usepackage{stmaryrd}
\usepackage{float}
\usepackage{textalpha}

\newcommand{\bs}[1]{{\boldsymbol{#1}}}

\newcommand{\bp}{\bs{p}}

\begin{document}
\title{From inverse-cascade to sub-diffusive dynamic scaling in driven disordered Bose fluids}
\author{Elisabeth Gliott}
\email{elisabeth.gliott@lkb.upmc.fr}
\affiliation{Laboratoire Kastler Brossel, Sorbonne Universit\'{e}, CNRS, ENS-PSL Research University, 
Coll\`{e}ge de France; 4 Place Jussieu, 75005 Paris, France}

\author{Adam Ran\c con}
\email{adam.rancon@univ-lille.fr}
\affiliation{Univ. Lille, CNRS, UMR 8523 -- PhLAM -- Laboratoire de Physique des Lasers Atomes et Mol\'ecules, F-59000 Lille, France}

\author{Nicolas Cherroret}
\email{nicolas.cherroret@lkb.upmc.fr}
\affiliation{Laboratoire Kastler Brossel, Sorbonne Universit\'{e}, CNRS, ENS-PSL Research University, 
Coll\`{e}ge de France; 4 Place Jussieu, 75005 Paris, France}

\begin{abstract}
We explore the emergence of universal dynamic scaling in an interacting Bose gas around the condensation transition, under the combined influence of an external driving force and spatial disorder. As time progresses, we find that the Bose gas crosses over three distinct dynamical regimes: (i) an inverse turbulent cascade where interactions dominate the drive, (ii) a stationary regime where the inverse cascade and the drive counterbalance one other, and (iii) a sub-diffusive cascade in energy space governed by the drive and disorder, a phenomenon recently observed experimentally. 
We show that all three dynamical regimes can be described by self-similar scaling laws.
\end{abstract}
\maketitle

In quantum many-body physics, the concept of universality is often linked with thermodynamic equilibrium near a phase transition. Universality, however, may also emerge far from equilibrium, where it acquires a special importance due to its 
association with predictive scaling laws that are largely independent of the microscopic, often intractable many-body dynamics \cite{Hohenberg1977, Eisert2015, Langen2015}. In systems undergoing quantum quenches, a key concept connected to universality is dynamic scaling \cite{Berges2008, Chantesana2019, Mikheev2019}, which describes a self-similar spatiotemporal evolution of correlations,  governed by universal exponents and scaling functions that are independent of initial conditions. Dynamic scaling was identified in a range of non-equilibrium phenomena such as coarsening \cite{Bray2002, Cugliandolo2015}, dynamical phase transitions \cite{Marino2022} or wave turbulence \cite{Zhu2023}. In practice, it often emerges by quenching a many-body system so that it crosses a symmetry-breaking phase transition, a protocol recently implemented in quantum-gas experiments \cite{Erne2018, Prufer2018, Bagnato2022, Abuzarli2022, Sunami2022, Gazo2023}.

An intriguing yet largely unexplored question is the robustness of universal dynamic scaling against external perturbations. Among those, periodic driving and disorder have been identified as key ingredients that significantly impact the relaxation of quantum many-body systems. Often combined with dissipation, driving can for instance lead to the formation of stationary turbulent spectra \cite{Nazarenko2011, Navon2019, Galka2022}. Disorder, on the other hand, was shown to trigger a variety of phenomena in non-equilibrium setups, such as weak and strong localization interference \cite{Jendrzejewski2012, Scoquart2020b, Cherroret2021}, prethermalization effects \cite{Larre2018, Scoquart2020a, Haldar2023, Scoquart2022} or the breakdown of thermalization through many-body localization \cite{Nandkishore2015, Laflorencie2018, Abanin2019}. 
Recently, by preparing a non-interacting Bose gas in the presence of \emph{both} a periodic driving force and spatial disorder, a cold-atom experiment highlighted a novel mechanism of sub-diffusion in energy space \cite{Martirosyan2024, Zhang2023}. At the same time, abruptly cooling an interacting Bose gas across the condensation transition in the same setup, but without any driving force, led to the observation of an inverse cascade \cite{Glidden2021}. In both experiments, the Bose gas evolution exhibited dynamic scaling.

\begin{figure}[t!]
\includegraphics[scale=0.7]{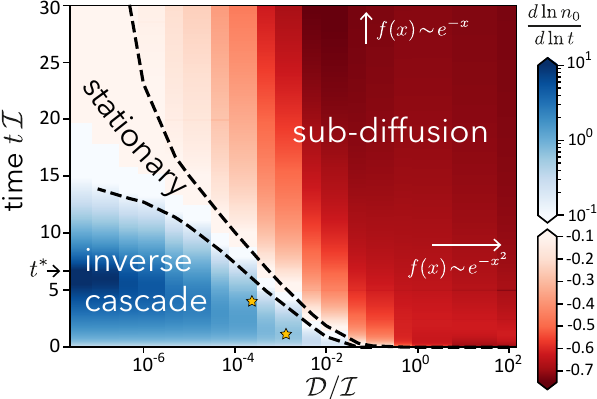}
\caption{\label{Fig:phasediag}
Dynamical phase diagram of an interacting Bose gas subjected to both a periodic force and disorder, parametrized by the parameter $\mathcal{D}$.
Colors measure the slope $\partial_{\ln t}\ln n_0(t)$ of the central energy density $n_0(t)$, which is respectively positive, near zero and negative in the inverse-cascade, stationary and sub-diffusive regions. These regions are separated by cross-overs, the dashed curves being guides pinpointing $|\partial_{\ln t}\ln n_0(t)|=0.1$.
Here $t$, $\mathcal{D}$ and $\mathcal{I}$ are in units of $\epsilon_0^{-1}$, $\epsilon_0^{3}$ and $\epsilon_0^{-1}$, respectively, where $\smash{\epsilon_0=(4\pi^2\rho_0)^{2/3}/(2m)}$.
Stars are experimental points of \cite{Martirosyan2024} operating near the stationary regime and corresponding to $(\mathcal{D}/\mathcal{I},t\mathcal{I})=(2\!\times\!10^{-4},3.9)$ and $(1.2\!\times\!10^{-3},0.6)$ \cite{Supplemental}. In the sub-diffusive region, the scaling function converges to a Gaussian (exponential) as $\mathcal{D}/\mathcal{I}$ ($t\mathcal{I}$) increases (as indicated by arrows).}
\end{figure}

These experimental findings naturally call for a clarification of the dynamical portrait resulting from the combined actions of interactions, drive, and disorder on Bose gases, and their relation to dynamic scaling. This is the task we undertake in this Letter
, by leveraging a quantum kinetic framework.
In the noninteracting limit, this approach enables us to recover the sub-diffusive dynamics observed in \cite{Martirosyan2024}. In the general case, we are able to construct the complete dynamical phase diagram of the system, which is given in Fig.~\ref{Fig:phasediag}. We find that as time progresses, the Bose gas crosses over three distinct regimes. Initially, interactions dominate the dynamics, leading to an inverse particle cascade associated with 
the dynamical formation of a condensate. At intermediate times, the Bose gas reaches a stationary regime where the inverse cascade and the external drive counterbalance one other. At late times, finally, the drive prevails, resulting in a  sub-diffusive direct energy cascade.
Remarkably, we find that all regimes are governed by self-similar evolutions of the energy distribution, with properties that nonetheless differ from the limits of vanishing drive or interactions.

We consider a three-dimensional degenerate Bose gas subjected to an external periodic force $\textbf{F}=F\cos(\omega t) \textbf{e}$ aligned along an arbitrary direction $ \textbf{e}$, and evolving in a weak, spatially disordered potential. We are interested in the quantum dynamics of the population $n_\bp(t)$ in a given momentum mode $\bp$, which we describe using the Boltzmann-type kinetic equation \cite{Griffin2009, Schwiete2013, Cherroret2015}
\begin{equation}
\label{Boltzmann_eq}
\partial_t n_\bp+\textbf{F}\cdot\partial_{\bp} n_\bp=C_{\text{d}}[n_\bp]+C_{\text{int}}[n_\bp],
\end{equation}
with the normalization $\int d^3\bp/(2\pi)^3 n_\bp(t)=\rho_0$, where $\rho_0$ is the mean gas density. 
The function $C_{\text{d}}[n_\bp]$ on the right-hand side is the rate of change of the population in mode $\bp$ due to elastic scattering of particles on the disorder. Assuming that scattering is isotropic and occurs at a rate $1/\tau_s$, we have $C_{\text{d}}[n_\bp]=[\langle n_\bp(t)\rangle_{\hat\bp}-n_\bp(t)]/\tau_s$, with $\langle\ldots\rangle_{\hat\bp}$ the average over the direction $\hat\bp$ of $\bp$. 
$C_{\text{int}}[n_\bp]$ is a collision integral describing repulsive  boson interactions \cite{Griffin2009}. Its expression is given below.
From now on, we focus on times $t\gg\tau_s$, where disorder scattering makes $n_\bp(t)$ nearly isotropic. In that case, we can use the spherical-harmonic expansion $n_\bp(t) \simeq \langle n_\bp(t)\rangle_{\hat\bp}+3\hat{\bp}\cdot\langle \hat{\bp}\, n_\bp(t)\rangle_{\hat{\bp}}$, and reduce Eq.~(\ref{Boltzmann_eq}) to a closed equation for the isotropic component $\langle n_\bp(t)\rangle_{\hat\bp}\equiv n_\epsilon(t)$, the energy distribution of the Bose gas (with $\epsilon=\bp^2/2m$ and $m$ the particle mass). By further averaging over the fast oscillations of the force, we obtain \cite{Cherroret2011, Supplemental}
\begin{equation}
\label{eq:kineticfinal}
\partial_t n_\epsilon=
\mathcal{D}
\Big(\partial^2_\epsilon+\frac{1}{2\epsilon}\partial_\epsilon\Big)  n_\epsilon+
C_{\text{int}}[n_\epsilon],
\end{equation}
with the collision integral
\begin{align}
&C_{\text{int}}[n_\epsilon]=\mathcal{I}\int_{\epsilon_3=\epsilon_1 + \epsilon_2-\epsilon>0}
\!\!\!\!\!\!\!\!\!\!\!\! \!\!\!\!\!\!\!\!\!d\epsilon_1 d \epsilon_2
\frac{\text{min}(\sqrt{\epsilon},\sqrt{\epsilon_1},\sqrt{\epsilon_2}, \sqrt{\epsilon_3})}{\sqrt{\epsilon}}\nonumber\\
&\times \left[n_{\epsilon_1}n_{\epsilon_2}(1+n_{\epsilon}+n_{\epsilon_3})-n_{\epsilon}n_{\epsilon_3}(1+n_{\epsilon_1}+n_{\epsilon_2})\right].
\end{align}
Equation (\ref{eq:kineticfinal}) has a single integral of motion, the total particle number: $\int\! d\epsilon\,\nu_\epsilon n_\epsilon(t)\!=\!\rho_0$, where $\nu_\epsilon\!=\! m^{3/2}\sqrt{\epsilon}/\sqrt{2}\pi^2$ is the density of states per unit volume (here and throughout the Letter we set $\hbar=1$).
The dynamics of $n_\epsilon(t)$ is governed by two core parameters. The first one, $\mathcal{D}=F^2D/2$ --with $D\propto \tau_s$ the diffusion coefficient--, controls the level of external drive and disorder. We take it energy independent, which corresponds to the scenario of \cite{Martirosyan2024}, see \cite{Zhang2023}. Notice that when $t\gg\tau_s$,  drive and disorder are coupled to one another. The second one,  $\mathcal{I}=g^2m^3/(2\pi^3)$, measures the strength of interactions. Equation 
\eqref{eq:kineticfinal} is the central tool of our work, which we use to investigate the full dynamics of the Bose gas. 
\begin{figure}
\includegraphics[scale=0.8]{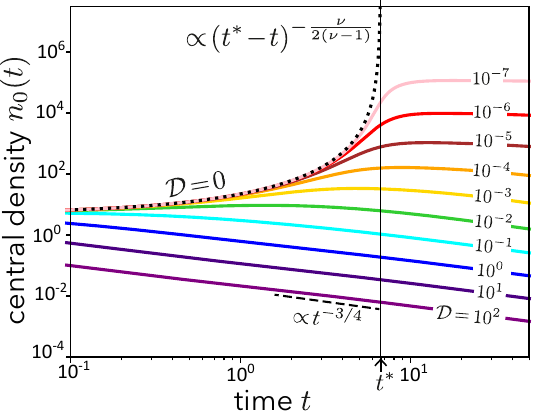}
\caption{\label{Fig:n0}
Central density $n_0(t)$ of the Bose gas vs. time for increasing values of $\mathcal{D}$, obtained by solving numerically Eq.~(\ref{eq:kineticfinal}).
At small $\mathcal{D}$, $n_0(t)$ first increases, indicating the presence of an inverse cascade. It then saturates and eventually decreases at longer times. The onset of this decay occurs earlier as $\mathcal{D}$ increases, and when $\mathcal{D}$ exceeds a certain threshold the inverse cascade no longer occurs. Note that at finite $\mathcal{D}$, the divergence of the isolated gas at $t=t^* \simeq 6.7$, Eq.~(\ref{eq:WTscaling}), is smoothed out.
Here $\mathcal{I}=1$, and units are the same as in Fig.~\ref{Fig:phasediag}.}
\end{figure}

It is instructive to first examine the limits of vanishing interactions and vanishing drive. For a noninteracting driven gas, i.e., $\mathcal{I}=0$ and $\mathcal{D}$ finite, Eq.~(\ref{eq:kineticfinal}) can be solved analytically, see \cite{Supplemental} for details. The long-time solution is independent of the  initial condition and given by
\begin{equation}
\label{eq:Gaussiansubdiff}
n_\epsilon(t)\propto {(4\mathcal{D}t)^{-3/4}} \exp[-\epsilon^2/(4\mathcal{D} t)],  \ \ \ \mathcal{D}\!\ne\!0,\, \mathcal I\!=\!0
\end{equation}
This corresponds to a direct \emph{sub-diffusive} cascade, interpreted as follows: due to the driving force, the momentum distribution first broadens along $\textbf{e}$, but this broadening is quickly redistributed in all directions through disorder scattering. The net effect is a random walk in energy space, as observed in the recent experiment \cite{Martirosyan2024}.
In the opposite limit of an isolated interacting gas, i.e., for $\mathcal{D}=0$ and $\mathcal{I}$ finite, Eq.~\eqref{eq:kineticfinal} reduces to the wave-turbulence equation, which was studied in \cite{Semikoz1995, Semikoz1997, Berloff2002, Josserand2006, Semisalov2021, Zhu2023}. 
An interesting property of this equation is that when the total energy of the initial distribution $n_\epsilon(0)$ is set sufficiently small, thus  modeling a cooling quench across the condensation transition,  $n_\epsilon(t)$ undergoes an inverse particle cascade of the form
\begin{equation}
\label{eq:WTscaling}
n_\epsilon(t)\propto (t^*\!-\!t)^{-{\frac{\nu}{2(\nu-1)}}} f[\epsilon/(t^*\!-\!t)^{\frac{1}{2(\nu-1)}}]  \ \ \ \mathcal{D}\!=\!0,\, I\!\ne\!0,
\end{equation}
 which describes particles flowing toward low energies. This flow is due to the dynamical formation of a Bose condensate as $t$ approaches the characteristic time $t^*$. The exponent $\nu\simeq 1.23$  governs the asymptotic decay $f(x\gg 1)\propto x^{-\nu}$ of the scaling function $f(x)$. For a smooth initial condition, one also has $\phi(0)=\text{constant}$, 
such that the central density $n_0(t)\propto (t^*\!-\!t)^{-{\frac{\nu}{2(\nu-1)}}}$ diverges algebraically near $t^*$.  This phenomenon is confirmed by a numerical calculation of $n_0(t)$ for  $\mathcal{D}=0$ presented in Fig.~\ref{Fig:n0} (dotted curve) for $\mathcal{I}=1$.

\begin{figure*}
\includegraphics[scale=0.45]{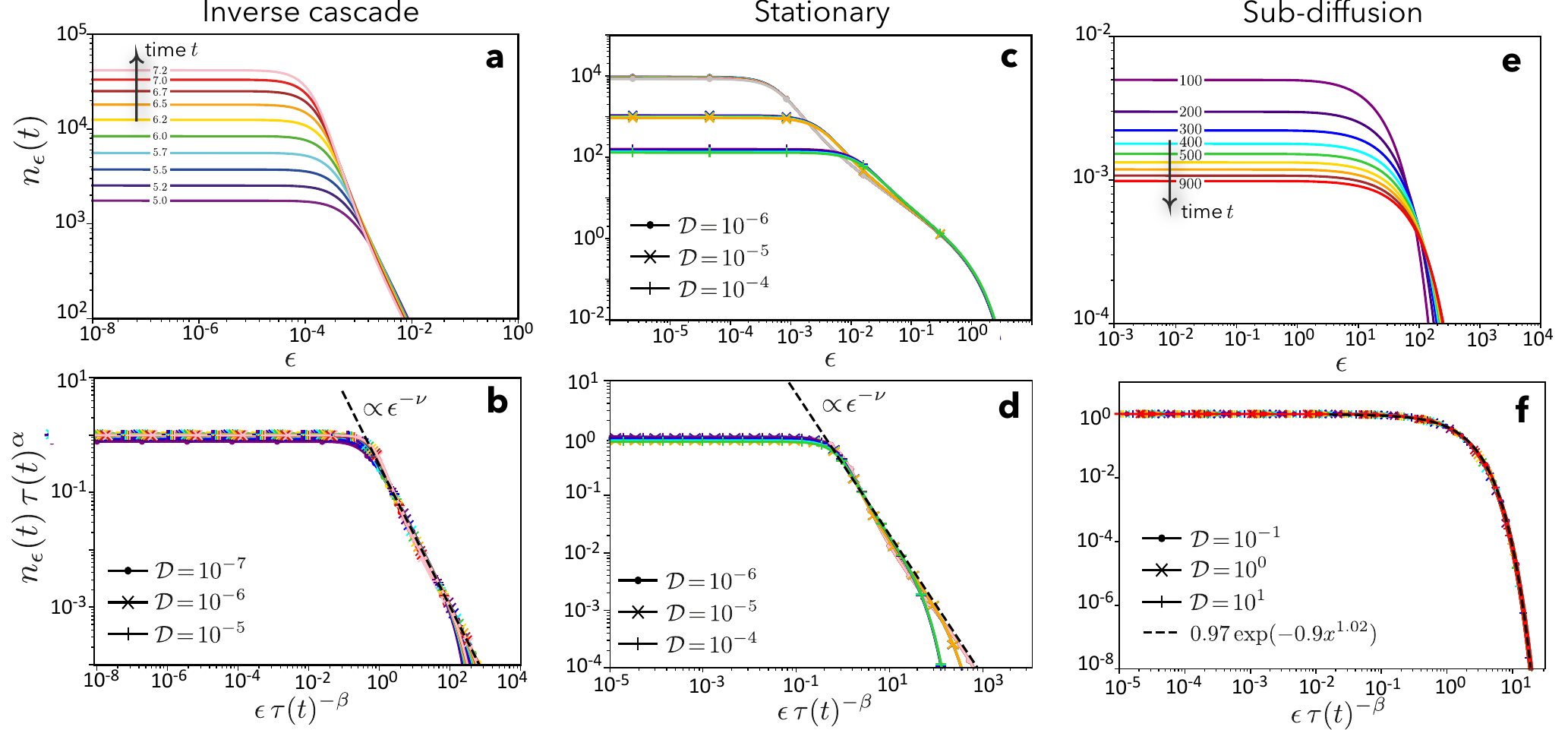}
\caption{\label{Fig:collapse}
Self-similar evolution of the energy distribution $n_\epsilon(t)$ in the three dynamical regimes. 
a) $n_\epsilon$ vs. time in the inverse-cascade region, at fixed $\mathcal{D}=10^{-7}$, and b) associated rescaling according to Eq.~(\ref{eq:self-similar}), using $\tau(t)\propto (t^*-t+\lambda t^\gamma)$ and $\smash{\alpha=\frac{\nu}{2(\nu-1)}}$, $\smash{\beta=\frac{1}{2(\nu-1)}}$ with $\nu\simeq1.23$. Notice that this rescaling includes a set of many $(\mathcal{D},t)$ values.
c) Distributions in the stationary region are time independent at fixed $\mathcal{D}$. d) Curves at different $\mathcal{D}$ are rescaled using Eq.~(\ref{eq:stationary}) with $\tau(t)=\tau_0=\text{constant}$, giving $\alpha/\beta\simeq1.23\simeq\nu$.
e) $n_\epsilon$ vs. time in the sub-diffusive region, at fixed $\mathcal{D}=10$, and f) dynamic rescaling for a set of $(\mathcal{D},t)$ values using $\tau(t) \propto \mathcal{D}t$.
Here $\mathcal{I}=1$, and units are the same as in Fig.~\ref{Fig:phasediag}.
}
\end{figure*}

Noticeably, in both limits $\mathcal{I}=0$ or $\mathcal{D}=0$, the energy distribution exhibits \emph{self-similar dynamic scaling}, i.e., it can be written in the generic form
 \begin{equation}
 \label{eq:self-similar}
n_\epsilon(t)= \tau(t)^{-\alpha} f\left[\epsilon\, \tau(t)^{-\beta}\right],
\end{equation}
with $\tau(t)\propto\mathcal{D}\,t$ and $\tau(t)\propto t^*\!-\!t$ for $\mathcal{I}=0$ and $\mathcal{D}=0$, respectively. In each limit, the exponents $\alpha$ and $\beta$ are universal, in the sense that they do not depend on $\mathcal{D}$ or $\mathcal{I}$, as well as on the details of the initial condition.
The natural question is whether such a property persists in the general scenario where \emph{both} $\mathcal{I}$ and $\mathcal{D}$ are finite, and how the system crosses over these two asymptotic scaling laws. To investigate this, we have performed extensive numerical simulations of the full kinetic equation \eqref{eq:kineticfinal}, by varying $\mathcal{D}$ at fixed $\mathcal{I}=1$, using an initial condition $n_\epsilon(t=0)$ such that an inverse cascade occurs in the limit $\mathcal{D}=0$ (see \cite{Supplemental} for details on the numerical implementation). 
Note that setting $\mathcal{I}=1$ does not imply any loss of generality; all subsequent results can be extended to $\mathcal{I}\ne 1$ by exploiting the invariance $\mathcal{D}\to \mathcal{D}/\mathcal{I}$ and $t\to t\times\mathcal{I}$ of Eq.~(\ref{eq:kineticfinal}).
We first show in Fig.~\ref{Fig:n0} the central energy density $n_0(t)$ as a function of time, for increasing values of $\mathcal{D}$. 
At small but finite $\mathcal{D}$, we observe that $n_0(t)$ initially grows and then saturates.
This saturation replaces the wave-turbulent divergence at $t^*$, which only occurs for $\mathcal{D}=0$. 
In other words, in the presence of external driving, the system continuously crosses over from an inverse cascade to a stationary regime at long times. 
This cross-over is displayed in the phase diagram of Fig.~\ref{Fig:phasediag}, which represents the local slope of $n_0(t)$ in the plane $(\mathcal{D},t)$. 
Physically, this stationary regime stems from the balance between the inverse cascade induced by interactions and the energy spreading caused by the combined effects of drive and disorder. 
This balance, however, is only transient: Fig.~\ref{Fig:n0}  suggests that at long time the external drive dominates, leading to a decay in $n_0(t)$. 
While this decay is clearly sub-diffusive at large $\mathcal{D}$, i.e., $n_0(t)\propto t^{-3/4}$, 
 we will argue below that this remains the case for any finite $\mathcal{D}$.
We observe, finally, that beyond a certain threshold in $\mathcal{D}$, neither the inverse cascade nor the stationary regime occurs.
From a set of $n_0(t)$ curves, we construct the diagram in Fig.~\ref{Fig:phasediag}, showcasing
different regions in the $(\mathcal{D},t)$ plane, referring to either an inverse cascade, a stationary regime, or a direct sub-diffusive cascade. This naturally invites us to investigate whether self-similar scaling laws like Eq.~\eqref{eq:self-similar} exist within these regions.

We show in Fig.~\ref{Fig:collapse}-a  distributions $n_\epsilon(t)$ as a function of time, for a fixed $\mathcal{D}=10^{-7}$ within the inverse-cascade regime. The flow of particles toward low energies is well visible and, moreover, we discern the persistence of an algebraic decay of $n_\epsilon(t)$  at intermediate energies, reminiscent of the 
limit \eqref{eq:WTscaling} of vanishing drive. Distributions at other values $\mathcal{D}=10^{-6}, 10^{-5}$ show a similar behavior, suggesting the existence of a dynamic scaling law in the inverse-cascade regime. This is confirmed by Fig.~\ref{Fig:collapse}-b, where we show three temporal data sets corresponding to different values of $\mathcal{D}$, all rescaled according to
 Eq.~(\ref{eq:self-similar}), with $\smash{\alpha=\frac{\nu}{2(\nu-1)}}$ and $\smash{\beta=\frac{1}{2(\nu-1)}}$, with $\nu\simeq1.23$. In this rescaling procedure, the non-universal features of the drive and the disorder are entirely absorbed in a renormalization of the effective time $\tau(t)\propto (t^*-t+\lambda t^\gamma)$ that parametrizes the scaling trajectory, through the parameters  $\lambda$ and $\gamma$ which depend on $\mathcal{D}$ \cite{footnote}. In particular, $\lambda=\lambda(\mathcal{D})\to {0}$ when $\mathcal{D}\to 0$, recovering Eq.~(\ref{eq:WTscaling}). In spirit, this renormalization is similar to the concept of pre-scaling recently put forward in \cite{Schmied2019, Gazo2023}.
 
We then show in Fig.~\ref{Fig:collapse}-c energy distributions computed within the stationary regime at different times and driving strengths.
Remarkably, for a given $\mathcal{D}$ the \emph{entire shape} of $n_\epsilon(t)$ is time independent. This can again be described by a dynamic scaling of the form of Eq.~(\ref{eq:self-similar}), where $\tau(t)=\tau_0$ is independent of time (though it depends on $\mathcal{D}$).
Consequently, the universal content of the dynamics reduces to a single exponent $\beta/\alpha$, with the individual values $\alpha$ and $\beta$ being arbitrary. This can be seen by defining  $n_0(\mathcal{D})=1/\tau_0^\alpha$ and rewriting Eq.~(\ref{eq:self-similar}) as
\begin{equation}
\label{eq:stationary}
n_\epsilon=n_0 f\Big(\epsilon n_0^{\beta/\alpha}\Big),
\end{equation} 
where the non-universal $\mathcal{D}$ dependence is encoded in $n_0$. A rescaling of the curves in Fig.~\ref{Fig:collapse}-d at different $\mathcal{D}$  with $\alpha/\beta\simeq 1.23\simeq \nu$ yields a collapse of distributions over more than 7 orders of magnitude in energy. Remarkably, we find once more that the scaling function obeys $f(x)\propto x^{-\nu}$ at large $x$.  In other words, despite the system operating in a regime significantly different from the non-driven limit $\mathcal{D}=0$, the many-body dynamics continues to display universal characteristics through $\nu$.

\begin{figure}[t!]
\includegraphics[scale=0.6]{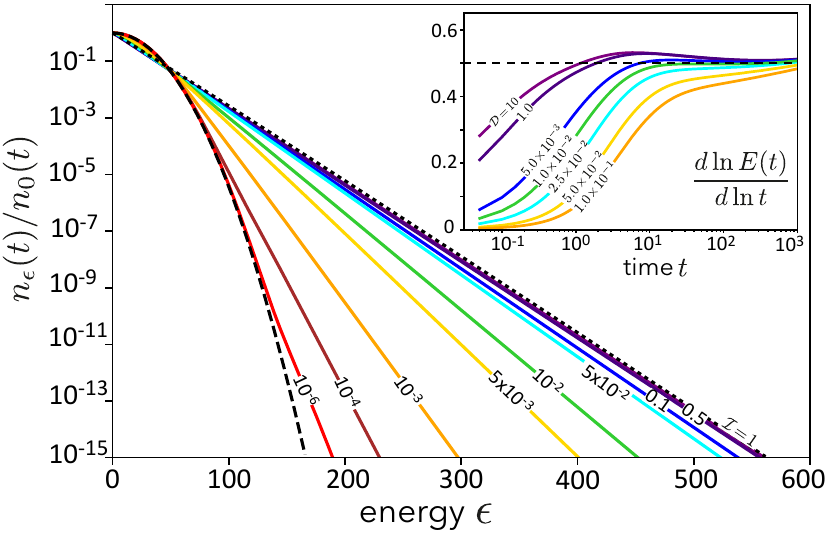}
\caption{\label{Fig:energy}
$n_\epsilon(t)$ vs. $\epsilon$ at fixed $t=200$ and $\mathcal{D}=1$, for decreasing values of $\mathcal{I}$. The distribution smoothly changes from  an exponential, Eq.~(\ref{eq:subdiffexp}), dotted curve, to a Gaussian, Eq.~(\ref{eq:Gaussiansubdiff}), dashed curve. 
Inset: logarithmic derivative of the total energy $E(t)$, demonstrating  $E(t)\propto \sqrt{t}$ for all $\mathcal{D}$ at long time.}
\end{figure}

The time evolution of $n_\epsilon(t)$ in the sub-diffusive region is finally displayed in Fig.~\ref{Fig:collapse}-e. Here, as the drive dominates over interactions, $n_\epsilon(t)$ exhibits a direct cascade toward high energies, reminiscent of the noninteracting law (\ref{eq:Gaussiansubdiff}). To verify whether this cascade is indeed sub-diffusive, in Fig.~\ref{Fig:collapse}-f we rescale $n_\epsilon(t)$ according to Eq.~(\ref{eq:self-similar}), by using the parametrization   $\tau(t) \propto \mathcal{D}t$ \cite{footnote} and including data sets at different $\mathcal{D}$ values spanning two orders of magnitude. We obtain a nearly perfect collapse of all data for $\alpha=0.74\simeq3/4$ and $\beta=0.5$, thus confirming a self-similar sub-diffusive dynamics. 
At first sight, this suggests that within the sub-diffusive regime, the impact of interactions is negligible, i.e., that the limit $\mathcal{I}=0$  is effectively recovered as soon as $\mathcal{D}/\mathcal{I}$ is large enough. This is not the case: as shown in Fig.~\ref{Fig:collapse}-f, in the sub-diffusive regime the scaling function $f(x)$ is very well captured by an exponential shape, in contrast to the Gaussian (\ref{eq:Gaussiansubdiff}) obtained for $\mathcal{I}=0$. Precisely, from our rescaling we find in good approximation
\begin{equation}
\label{eq:subdiffexp}
n_\epsilon(t\!\to\!\infty)\propto\! \frac{1}{(\mathcal{D}t)^{3/4}}\exp\left[-\frac{\epsilon }{(3\mathcal{D}t/2)^{1/2}}\right], \ \mathcal{D}, \mathcal I\!\ne\!0.
\end{equation}
In other words, residual interactions, even small, modify the sub-diffusive scaling function at long time. 
The natural question that arises is how the non-interacting limit (\ref{eq:Gaussiansubdiff}) is recovered from Eq.~(\ref{eq:subdiffexp}) as $\mathcal{I}\to0$. 
This is addressed in Fig.~\ref{Fig:energy}, which shows $n_\epsilon(t)$ computed numerically in the sub-diffusive region at a fixed (long) time for different $\mathcal{I}$: As interactions are reduced, the scaling function smoothly changes from Eq.~(\ref{eq:subdiffexp}) to Eq.~(\ref{eq:Gaussiansubdiff}). The
dynamical evolution of the scaling function toward either shape is indicated in Fig.~\ref{Fig:phasediag}.

While the kinetic equation conserves particle number, the total energy per particle $E(t)=\rho_0^{-1}\int d\epsilon\,\epsilon\, \nu_\epsilon n_\epsilon(t)$ varies throughout the evolution. A non-trivial question we finally address is the long-time limit of $E(t)$, where one may ask whether sub-diffusion always prevails even for small $\mathcal{D}$.
 To clarify this, we have computed $E(t)$ from Eq.~(\ref{eq:kineticfinal}) up to the largest possible times. We show in the inset of Fig.~\ref{Fig:energy} the logarithmic derivative of $E(t)$ for different values of $\mathcal{D}$ spanning four orders of magnitude. We find that $\partial_{\ln t}\ln E(t) \to 1/2$ in all cases, confirming the sub-diffusive scaling $E(t\to\infty)\propto \sqrt{t}$ for all $\mathcal{D}$. A further analysis presented in \cite{Supplemental} shows that
\begin{equation}
\label{eq:energy}
E(t\to\infty)\simeq 1.18 E_0(t)\propto \sqrt{\mathcal{D}\,t},
\end{equation}
with $E_0(t)$ the total energy for $\mathcal{I}=0$. This law, which we conjecture valid for any finite $\mathcal{I}$, is fully compatible with the invariance $(\mathcal{D},t)\to (\mathcal{D}/\mathcal{I},t\times \mathcal{I})$ of Eq.~(\ref{eq:kineticfinal}). In turn, the net impact of interactions at long times lies in the numerical prefactor $1.18>1$.
In particular, in the sub-diffusive regime, this prefactor encapsulates the change of shape of the scaling function from Eq.~(\ref{eq:Gaussiansubdiff}) to (\ref{eq:subdiffexp}). Indeed, taking for the scaling function $f(x)$ (for $\mathcal{I}\ne0$) and $f_0(x)$ (for $\mathcal{I}=0$) respectively the exponential and Gaussian laws (\ref{eq:subdiffexp}) and (\ref{eq:Gaussiansubdiff}), we find $E(t)/E_0(t)=\int dx x^{3/2}f(x)/\int dx x^{3/2}f_0(x)\simeq 1.25$, close to 1.18 \cite{footnote2}.

In this work, we have established the dynamical portrait of driven disordered Bose gases near their condensation transition, connecting scenarios considered in recent cold-atom experiments. In particular, we have identified a novel form of stationary dynamics resulting from the competition between driving and interactions and exhibiting universal scaling. In \cite{Martirosyan2024}, measurements of momentum distributions carried out in a $^{39}\text{K}$ gas of $10^5$ atoms with scattering lengths $a=20a_0$ and $50a_0$ (with $a_0$ the Bohr radius) up to $t\simeq 1\,\text{s}$ were falling near this regime, as shown by the star symbols in Fig.~\ref{Fig:phasediag} \cite{Supplemental}. 
This offers exciting perspectives for future explorations of this stationary regime or of more complex scenarios involving, e.g., localization phenomena at stronger disorder or correlated gases.

NC and EG thank helpful discussions with Cl\'ement Duval and Dominique Delande. This work has benefited from the financial support of Agence Nationale de la Recherche (ANR), France, under the Grant  No.~ANR-19-CE30-0028-01 CONFOCAL. AR was partially supported by an IEA CNRS project and by the “PHC COGITO” program (project number: 49149VE), funded by the French Ministry for Europe and Foreign Affairs, the French Ministry for Higher Education and Research, and The Croatian Ministry of Science and Education.

\pagebreak
 \onecolumngrid
\begin{center}
\newpage\textbf{\Large Supplemental Material}
\end{center}

\setcounter{equation}{0}
\setcounter{figure}{0}
\setcounter{table}{0}
\setcounter{page}{1}
\makeatletter
\renewcommand{\theequation}{S\arabic{equation}}
\renewcommand{\thefigure}{S\arabic{figure}}
\renewcommand{\bibnumfmt}[1]{[S#1]}
\renewcommand{\citenumfont}[1]{S#1}

In this supplemental, we derive the kinetic equation (2) of the main text and its solution in the absence of interactions. We also describe the numerical method employed to solve it in the general case, and provide details about the rescaling procedure of energy distributions in the various dynamical regimes. Finally, we briefly explain how experimental data of \cite{Martirosyan2024} are exploited.
In the following, we set $\hbar=1$.

\section{Derivation of the kinetic equation}
\label{Sec:der}

In this section, we derive Eq.~(2) of the main text. Our starting point is the Boltzmann equation
\begin{equation}
\label{Boltzmann_eq}
\partial_t n_\bp+\textbf{F}\cdot\partial_{\bp} n_\bp=
\frac{\langle n_\bp\rangle_{\hat\bp}-n_\bp}{\tau_s}+C_{\text{int}}[n_\bp],
\end{equation}
for the distribution function $n_\bp(t)$. The first term on the right-hand side represents the elastic scattering caused by the disordered potential (with scattering rate $\tau_s^{-1}$), whereas $C_{\text{int}}[n_\bp]$ accounts for particle collisions. For a weakly interacting Bose gas in three dimensions, we have \cite{Griffin2009, Schwiete2013}
\begin{equation}
     C_\text{int}[n_\bp] = 4\pi g^2 \int \frac{d\bp_1 d\bp_2 d\bp_3}{(2\pi)^{6}} \delta(\bp_3\! +\! \bp - \bp_1\! -\!\bp_2) \delta(\epsilon_3\! +\! \epsilon\! -\!\epsilon_2\!-\!\epsilon_1)
    [ n_{\bp_2} n_{\bp_1} (1\!+\!n_\bp\!+\!n_{\bp_3})\! -\!  n_{\bp} n_{\bp_3} (1\!+\!n_{\bp_2}\!+\!n_{\bp_1}) ],
\end{equation}
with $g=4\pi a/m$ the interaction strength ($a$ is the s-wave scattering length and $m$ the atomic mass).
In our work, we focus on the hydrodynamic regime of long times where multiple scattering on the disorder has nearly isotropized the distribution $n_{\bp}(t)$. We also assume that scattering events on the disorder, of time scale $\tau_s$, occur more frequently than inter-particle collisions, for which the characteristic time scale is denoted by $\tau_\text{int}$. In other words, we restrict ourselves to a regime where
\begin{equation}
\label{eq:hydro}
t, \tau_\text{int}\gg \tau_s.
\end{equation}
Since the distribution $n_\bp(t)$ is nearly isotropic in this regime, its zeroth angular harmonic is the dominant one and we can write:
\begin{equation}
n_\bp(t)\simeq  n_p(t)+3\, \hat{\bp}\cdot \textbf{n}_p(t)
\end{equation}
where $\hat{\bp} \equiv \bp/p$, $n_p(t) \equiv \langle n_\bp(t) \rangle_{\hat{\bp}} $ and $\textbf{n}_p(t)\equiv \langle \hat{\bp}\,n_\bp(t)\rangle_{\hat{\bp} }$, where $\langle\ldots\rangle_{\hat{\bp}}$ refers to an average over the direction of $\bp$. By definition, this expansion supposes $|\textbf{n}_p(t)|\ll n_p(t)$.
Inserting it in Eq.~(1) and projecting onto the zeroth and first angular harmonics, we obtain the coupled relations
\begin{align}
\label{eqf0}
&\partial_t n_p(t)+v \textbf{F}\cdot\partial_\epsilon \,\textbf{n}_p(t)+
\frac{2}{p}\textbf{F}\cdot \textbf{n}_p(t)=\langle C_\text{int}[n_\bp]\rangle_{\hat{\bp}}\\
\label{eqf1}
&\partial_t \textbf{n}_p(t)+\frac{v}{d}\textbf{F}\partial_\epsilon\, n_p(t)=-\frac{\textbf{n}_p(t)}{\tau_s}+\langle \hat{\bp}\, C_\text{int}[n_\bp]\rangle_{\hat{\bp}}
\end{align}
where $\epsilon\equiv \bp^2/(2m)$ and $v\equiv p/m$. At long times $t\gg\tau_s$, we have $|\partial_t \textbf{n}_p(t)|\ll |\textbf{n}_p(t)|/\tau_s$, so that the first term in the left-hand side of Eq.~(\ref{eqf1}) can be neglected. Furthermore, we have 
\begin{equation}
\langle \hat{\bp}\, C_\text{int}[n_\bp]\rangle_{\hat{\bp}}=\frac{\langle \hat{\bp}\, C_\text{int}[n_\bp]\rangle_{\hat{\bp}}}{\langle C_\text{int}[n_\bp]\rangle_{\hat{\bp}}}\times
\langle  C_\text{int}[n_\bp]\rangle_{\hat{\bp}}
\sim \frac{|\textbf{n}_p(t)|}{n_p(t)}\times\frac{n_p(t)}{\tau_\text{int}}\sim \frac{|\textbf{n}_p(t)|}{\tau_\text{int}}\ll  \frac{|\textbf{n}_p(t)|}{\tau_s}.
\end{equation}
Therefore, the second term in the right-hand side of Eq.~(\ref{eqf1}) can be neglected as well. 
In turn, Eq.~(\ref{eqf1}) yields $\textbf{n}_p(t)\simeq -\tau_s \frac{v}{d}\textbf{F}\partial_\epsilon n_p(t)$, which once substituted in Eq.~(\ref{eqf0}) leads to a closed equation for $n_p(t)$ \cite{Cherroret2011, Schwiete2013}:
\begin{equation}
\label{eq:kineticgeneral}
\partial_t n_p(t) -\left[\textbf{F}(t)\partial_\epsilon+\frac{\textbf{F}(t)}{2\epsilon}\right] D \left[\textbf{F}(t) \partial_\epsilon\right] n_p(t)=\langle C_\text{int}[n_\bp]\rangle_{\hat{\bp}},
\end{equation}
where we introduced the diffusion coefficient $D\equiv v^2\tau_s/3$. Let us finally mention two additional assumptions. First, we  assume that the oscillations of the force $\mathbf{F} = F\cos(\omega t)\,\mathbf{e}$ occur at a fast scale $\omega^{-1}\sim \tau$, such that they can be averaged out in the hydrodynamic description. Second, we take the diffusion coefficient $D$ to be constant, which corresponds to the scenario of \cite{Martirosyan2024}. With this, Eq.~(\ref{eq:kineticgeneral}) reduces to 
\begin{equation}
\label{eq:kin}
    \partial_t n_\epsilon = \mathcal{D} \biggl(\partial_\epsilon^2  + \frac{1}{2\epsilon} \partial_\epsilon \biggr) n_\epsilon + \langle C_\text{int}[n_\bp]\rangle_{\hat{\bp}},
\end{equation}
with $\mathcal{D}=D F^2/2$.
This is essentially Eq.~(2) of the main text, upon renaming $n_\epsilon(t)=n_p(t)$ with $\epsilon = p^2/2m$. At leading order, the angularly averaged collision integral in the right-hand side reads
\begin{align}
\langle C_\text{int}[n_\bp]\rangle_{\hat{\bp}}\simeq
4\pi g^2\!\! \int \frac{d\hat{\bp}\,d\bp_1 d\bp_2 d\bp_3}{(2\pi)^{6}} \delta(\bp_3\! +\! \bp - \bp_1\! -\!\bp_2) \delta(\epsilon_3\! +\! \epsilon\! -\!\epsilon_2\!-\!\epsilon_1)
    [ n_{p_2} n_{p_1} (1\!+\!n_p\!+\!n_{p_3})\! -\!  n_{p} n_{p_3} (1\!+\!n_{p_2}\!+\!n_{p_1})].\nonumber
\end{align}
By computing the angular averages $\int d\hat{\bp}\,d\hat{\bp}_1\,d\hat{\bp}_2\,d\hat{\bp}_3$, which only involve the first Dirac function \cite{Schwiete2013}, we finally obtain $C_{\text{int}}[n_\epsilon]$ in Eq.~(3) of the main text. 

\section{Analytical solution in the absence of interactions}
\label{Sec:sol}

In this section, we derive the general solution of the kinetic equation  (\ref{eq:kin}) for $\mathcal{I}=0$, as well as its long-time limit, Eq.~(4) of the main text. Here and in the following, we express energies $\epsilon$,  time $t$, driving $\mathcal{D}$ and interaction strength $\mathcal{I}$ respectively in units of $\epsilon_0$, $\epsilon_0^{-1}$, $\epsilon_0^{3}$ and $\epsilon_0^{-1}$, where $\smash{\epsilon_0=(4\pi^2\rho_0)^{2/3}/(2m)}$.
For $\mathcal{I} = 0$, Eq.~(\ref{eq:kin}) reads
\begin{equation}
\partial_t n_\epsilon(t)  = \mathcal{D}\mathcal{L} n_\epsilon(t),
\label{eq_diff}
\end{equation}
where the operator $\mathcal{L} =\sqrt{\epsilon}^{-1}\partial_\epsilon (\sqrt{\epsilon}\partial_\epsilon \cdot)$ is semi-definite, negative, and symmetric with respect to the scalar product $\langle \cdot | \cdot\rangle$ defined as $\
\langle f | g\rangle=\int_0^\infty d\epsilon \, w(\epsilon) f(\epsilon)g(\epsilon)$ with weight $w(\epsilon)=\sqrt{\epsilon}$. We impose that the functional space corresponds to analytic functions at the origin, decaying at infinity. 
The eigenfunctions $\phi_\lambda(\epsilon)$ of $\mathcal{L}$ are obtained from standard Sturm-Liouville theory. They can be labeled with a continuous index $\lambda\in[0,\infty[$ and satisfy  $\mathcal{L}\phi_\lambda=-\lambda^2\phi_\lambda$,
with $\langle \phi_\lambda | \phi_{\lambda'}\rangle = \delta(\lambda-\lambda')$ and $\int_0^\infty d\lambda \phi_\lambda(\epsilon)\phi_\lambda(\epsilon') = {\delta(\epsilon-\epsilon')}/w(\epsilon)$. We find  $\smash{\phi_\lambda(\epsilon ) = \sqrt{\lambda} \epsilon^{1/4}J_{-1/4}(\lambda\epsilon)}$ with $J_\nu$ are  Bessel functions of the first kind. 

Given an initial condition $n_\epsilon(0)$, the solution of Eq.~(\ref{eq_diff}) is $n_\epsilon(t) = \int_0^{+\infty} d\epsilon' G_t(\epsilon,\epsilon') w(\epsilon') n_{\epsilon'}(0)$, where the Green's function $ G_t(\epsilon,\epsilon')$ satisfies $\partial_t G_t(\epsilon,\epsilon') = \mathcal{D} \mathcal{L}G_t(\epsilon,\epsilon')$ and $G_0(\epsilon,\epsilon') = {\delta(\epsilon - \epsilon')}/w(\epsilon)$. Its spectral decomposition is
\begin{equation}
G_t(\epsilon,\epsilon') =\int_0^\infty d\lambda\, e^{-\lambda^2 \mathcal{D}t} \phi_\lambda(\epsilon)\phi_\lambda(\epsilon').
\end{equation}
Using the above expression of eigenfunctions and  the identity 10.22.67 of  \href{https://dlmf.nist.gov/10.22}{DLMF}, we obtain 
\begin{equation}
\begin{split}
G_t(\epsilon,\epsilon') = \frac{(\epsilon\epsilon')^{\frac14}}{2\mathcal{D}t} e^{-\frac{\epsilon^2+\epsilon'^2}{4\mathcal{D}t}}I_{-\frac14}\left(\frac{\epsilon\epsilon'}{2\mathcal{D}t}\right),
\end{split}
\end{equation}
where $I_{\nu}(z)$ are modified Bessel functions of the first kind. We infer
\begin{equation}\label{arbitrary_solution}
    n_\epsilon(t) = \frac{\epsilon^{1/4}}{2\mathcal{D}t} e^{-\epsilon^2/4\mathcal{D}t} \int_0^{+\infty} d\epsilon' \epsilon'^{3/4} e^{-\epsilon'^2/4\mathcal{D}t} I_{-1/4}\biggl(\frac{\epsilon \epsilon'}{2\mathcal{D}t}\biggr) n_{\epsilon'}(0).
\end{equation}
At long times, we can expand the Bessel function using $I_{-1/4}(z/2) \simeq \sqrt{2}/[z^{1/4}\Gamma(3/4)]$. By further using the normalization condition $\smash{\int_0^{+\infty} d\epsilon' \sqrt{\epsilon'} n_0(\epsilon')=1}$, we deduce
\begin{equation}
    n_\epsilon(t) \simeq \frac{e^{-\epsilon^2/4\mathcal{D}t}}{\sqrt{2}\Gamma(3/4)(\mathcal{D}t)^{3/4}},
\end{equation}
which is Eq.~(4) of the main text.

\section{Numerical resolution of the kinetic equation}
\label{Sec:num}

\subsection{Numerical procedure}

In our numerical simulations, we solve the kinetic equation
\begin{equation}
    \partial_t n_\epsilon(t) = (A_I + A_D) n_\epsilon(t),
\end{equation}
where $A_I$ corresponds to the collision integral (3) of the main text and $A_D = \mathcal{D}(\partial^2_\epsilon + \frac{1}{2\epsilon}\partial_\epsilon )$. 
To achieve this goal across the wide range of parameters covered by the phase diagram in Fig.~1 of the main text, a finite difference method proved inadequate so we instead use a second-order Strang operator-splitting method \cite{Spohn_strang, Linge}, by decomposing one physical time-step $dt$ into into three parts: we first let $n_\epsilon(t)$ evolve under $A_D$ for a time-step $dt/2$, then under $A_I$ for $dt$, and finally again under $A_D$ for $dt/2$. The global error for this type of integration is of the order $O(dt^2)$ \cite{Linge}. In our simulations, we have  used $dt=0.01$ for simulations typically running  up to $t=50$, and $dt=0.1$ for simulations running up to $t=1000$.

The evolution under $A_D$ during $dt$ can be calculated using Eq.~(\ref{arbitrary_solution}) to calculate the energy distribution at $t + dt$ from the distribution at $t$:
\begin{equation}
    n_\epsilon(t + dt) = \frac{\epsilon^{1/4}}{2\mathcal{D}dt} \int_0^{+\infty} d\epsilon' \epsilon'^{3/4} e^{-(\epsilon - \epsilon')^2/4\mathcal{D}dt}
    I_{-1/4}\biggl(\frac{\epsilon \epsilon'}{2\mathcal{D}dt}\biggr) e^{\epsilon\epsilon'/2\mathcal{D}dt}
    n_{\epsilon'}(t).
\end{equation}
Since the integrand is typically peaked around $\epsilon'=\epsilon$, to achieve a high enough precision in the computation of the integral we replace the bounds from $0$ to $\infty$ by a definite interval $[\epsilon_1',\epsilon_2']$, where $\epsilon_1'$ and $\epsilon_2'$ are chosen as the points where $\smash{e^{-(\epsilon - \epsilon_{1,2}')^2/4\mathcal{D}dt} = 10^{-25}}$.

The evolution under the nonlinear operator  $A_I$ is calculated using the fourth-order Runge-Kutta method. The two-dimensional integral over $\epsilon_1$ and $\epsilon_2$ in the collision integral [Eq.~(3) of the main text] is divided in four regions corresponding to the four possible values taken by $W(\epsilon,\epsilon_1,\epsilon_2) = \text{min}(\sqrt{\epsilon},\sqrt{\epsilon_1},\sqrt{\epsilon_2}, \sqrt{\epsilon_3})/\sqrt{\epsilon}$ with $\epsilon_3=\epsilon_1+\epsilon_2-\epsilon$.
The symmetry along the $\epsilon_1 = \epsilon_2$ axis then allows us to reduce these four integration domains to three, with the collision integral given by the sum
\begin{equation}
    C_{\text{int}}[n_\epsilon] = 
    2\int_{\epsilon/2}^{\epsilon} d\epsilon_1 \int_{\epsilon - \epsilon_1}^{\epsilon_1} d\epsilon_2 \sqrt{\epsilon_3/\epsilon} \text{ occ}[n_\epsilon]
    +\int_{\epsilon}^{\epsilon_{\text{max}}} d\epsilon_1 \int_{\epsilon}^{\epsilon_{\text{max}}} d\epsilon_2 \text{ occ}[n_\epsilon]
    +2\int_{\epsilon}^{\epsilon_{\text{max}}} d\epsilon_1 \int_{\epsilon_{\text{min}}}^{\epsilon} d\epsilon_2 \sqrt{\epsilon_1/\epsilon}\text{ occ}[n_\epsilon],
\end{equation}
where $\text{ occ}[n_\epsilon] = [n_{\epsilon_1}n_{\epsilon_2}(1+n_{\epsilon}+n_{\epsilon_3})-n_{\epsilon}n_{\epsilon_3}(1+n_{\epsilon_1}+n_{\epsilon_2})]$. 
The precision of our numerical results can be estimated from norm conservation. Typically, in our simulations norm conservation is satisfied at $10^{-4}$ up to $t=50$, and at $10^{-2}$ up to $t=1000$. 

All  integrals involved in the procedure described above  are calculated using the tanh-sinh quadrature method \cite{TAKAHASI} with a precision of order $4$. 
This integration technique allows us to obtain very precise results with relatively few integration points (in practice $\sim100$). Furthermore, at each step of the time integration we interpolate the energy distribution using a monotonic 1D cubic interpolation method 
\cite{Pchip} with $2000$ interpolation points.

\subsection{Initial condition}

In our numerical simulations, we use the following Gaussian initial condition (in dimensionless units):
\begin{equation}\label{IC}
    n_\epsilon(t = 0) = \frac{2}{\Gamma(3/4)\epsilon_c^{3/2}} e^{-(\epsilon/\epsilon_c)^2}.
\end{equation}
This distribution is normalized according to $\int_0^\infty d\epsilon \sqrt{\epsilon}\, n_\epsilon(t) = 1$. In our work, we choose the width $\epsilon_c$ of the Gaussian small enough so that an inverse cascade associated with the formation of a condensate occurs in the absence of disorder and external driving. Such a condensate appears when the constraints of particle and energy conservation become incompatible with the evolution of the distribution  (\ref{IC}) under Eq.~(2) of the main text (with $\mathcal{D}=0$) toward a pure thermal distribution $[e^{(\epsilon - \mu)/T}-1]^{-1}$, where $\mu$ and $T$ are respectively the chemical potential and the temperature at equilibrium. These constraints read
\begin{align}
\int_0^{+\infty} d\epsilon \sqrt{\epsilon} n_\epsilon(t = 0) &= \int_0^{+\infty} d\epsilon\, \frac{\sqrt{\epsilon}}{e^{(\epsilon-\mu)/T}-1}, \\
 \int_0^{+\infty} d\epsilon \epsilon^{3/2} n_\epsilon(t = 0) &= \int_0^{+\infty} d\epsilon \, \frac{\epsilon^{3/2}}{e^{(\epsilon-\mu)/T}-1}.
\end{align}
Performing the integrals and combining these two equations, we obtain a relation between $\epsilon_c$ and the ratio $\mu/T$:
\begin{equation}\label{eq_CI}
    \frac{\Gamma(5/4)}{\Gamma(3/4)}\,\epsilon_c = \frac{6}{\pi} \frac{\text{Li}_{5/2}(e^{\mu/T})}{[\text{Li}_{3/2}(e^{\mu/T})]^3},
\end{equation}
where $\text{Li}_s(x)$ is the polylogarithm function of order $s$. At the onset of Bose-Einstein condensation, the chemical potential $\mu$ vanishes, corresponding to
\begin{equation}
    \epsilon_c(\mu = 0) = \frac{6}{\pi} \frac{\Gamma(3/4)}{\Gamma(5/4)}  \frac{\zeta(5/2)}{\zeta(3/2)^3} \simeq 1.92.
\end{equation}
According to Eq.~(\ref{eq_CI}), $\epsilon_c$ decreases as $\mu$ increases. The chemical potential being strictly negative, Eq.~\ref{eq_CI} has therefore no solution when $\epsilon_c < \epsilon_c(\mu = 0)$. This is the condition for the inverse cascade to occur. In our study, we choose $\epsilon_c = 1/\sqrt{5}\simeq 0.447$.

\section{Rescaling procedure and total energy}
\label{Sec:resc}

In this section, we provide details on the procedure used to rescale the energy distributions in the sub-diffusive, stationary and inverse-cascade regimes using the self-similar form
\begin{equation}
    n_\epsilon(t) = 
   \tau(t)^{-\alpha} f[\epsilon/\tau(t)^{\beta}].
\end{equation}
In the three regimes, we arbitrarily impose  $f(0)=1$, which amounts to fixing the numerical prefactor in $\tau(t)$ so that $n_0(t)\tau(t)^\alpha=1$.

\subsection{Sub-diffusive regime}

\begin{figure}[H]
\centering
\includegraphics[scale=0.6]{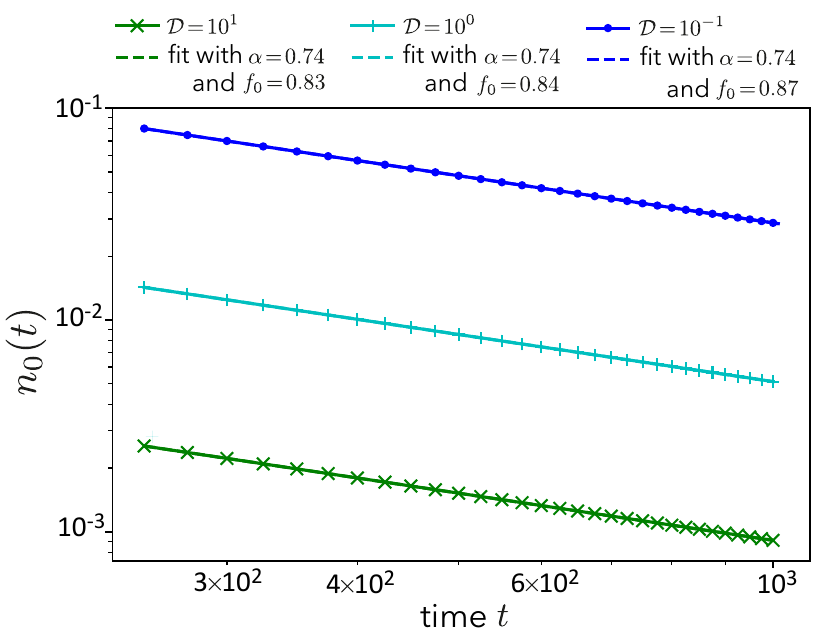}
\includegraphics[scale=0.6]{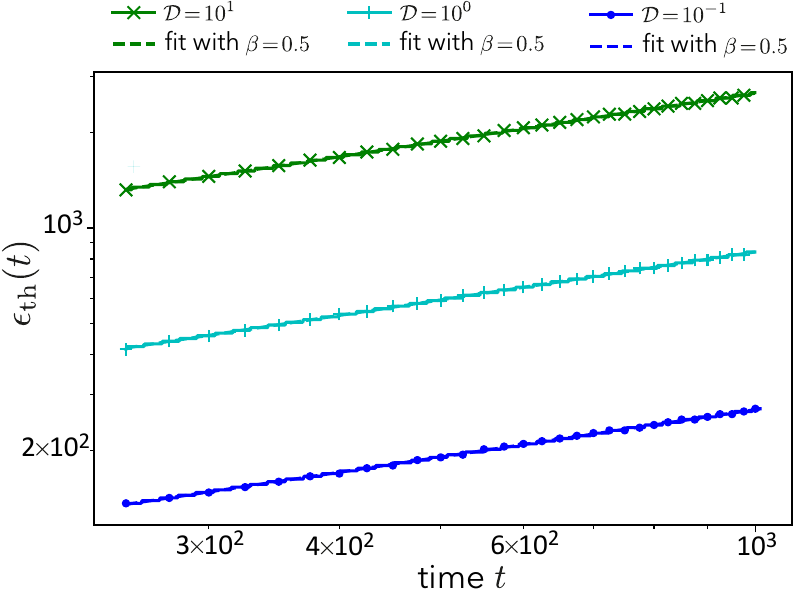}
\caption{
Left: Fits of the central density $n_0(t)$ to the function $f_0(\mathcal{D})\times (\mathcal{D}t)^{-\alpha}$ for different $\mathcal{D}$ in the sub-diffusive regime. We find $\alpha \simeq 0.74$ and $f_0 \simeq 0.85\pm 0.2$ for the fit parameters, suggesting that both are in good approximation independent of $\mathcal{D}$.
Right: Fits of $\epsilon_{\text{th}}(t)$ to the function $c_0(\mathcal{D}t)^{\beta}$, for the same three values of $\mathcal{D}$. Here $\epsilon_{\text{th}}(t)$ is defined as the energy at which $n_\epsilon(t)/n_0(t) = 10^{-12}$. The fit provides $\beta \simeq 0.50$, again independent of $\mathcal{D}$. In both figures, fits are performed within the temporal interval $[250,1000]$.
}
\label{Fig:fit_subdiffusive}
\end{figure}

To obtain the rescaled distribution shown in Fig.~3-f of the main text, we proceed in two steps where we successively determine $\alpha$ and $\beta$. First, we fit the central density $n_0(t)$ to a function $\tau(t)^{-\alpha}$, where we choose $\tau(t) = f_0(\mathcal{D})^{-1/\alpha} \mathcal{D}t$, with $f_0(\mathcal{D})$ and $\alpha$ fit parameters. Such fits are shown in the left plot of Fig.~\ref{Fig:fit_subdiffusive} for different values of $\mathcal{D}$. They give $\alpha=0.74$ and $f_0(\mathcal{D})=0.85$, both essentially independent of $\mathcal{D}$. 
In the second step, we determine $\beta$ from the behavior of the energy distribution at large energies.
To do so, we study the energy abscissa $\epsilon_{\text{th}}(t)$ at which $n_{\epsilon_{\text{th}}}(t)/n_0(t) = 10^{-12}$  as a function of time. We then fit this quantity at long times with $c_0 (\mathcal{D}t)^{\beta}$ to extract $\beta$. 
Such fits are shown in the right plot of Fig.~\ref{Fig:fit_subdiffusive} for different values of $\mathcal{D}$. They give $\beta=0.50$, again independent of $\mathcal{D}$. 

In Fig.~3-f of the main text, we have fitted the scaling function $f(x)$ resulting from the collapse of the set of $(\mathcal{D},t)$ curves in the sub-diffusive regime to $f_{\text{fit}}(x) =f_0 e^{-ax^{\kappa}}$, with $f_0$, $a$, and $\kappa$ fit parameters. This fit is done in the interval of energy $\epsilon$ $[10^{-2},10^{5}]$ to better capture the rapidly decreasing function at high energies, and yields $f_0 = 0.97$, $a=0.9$ and $\kappa=1.02$, for which the norm is close to 1 : $\smash{\int_0^{+\infty} dx\sqrt{x}f_{\text{fit}}(x) \simeq 0.98}$.  

\subsection{Inverse cascade}

\begin{figure}[h]
\centering
\includegraphics[scale=0.7]{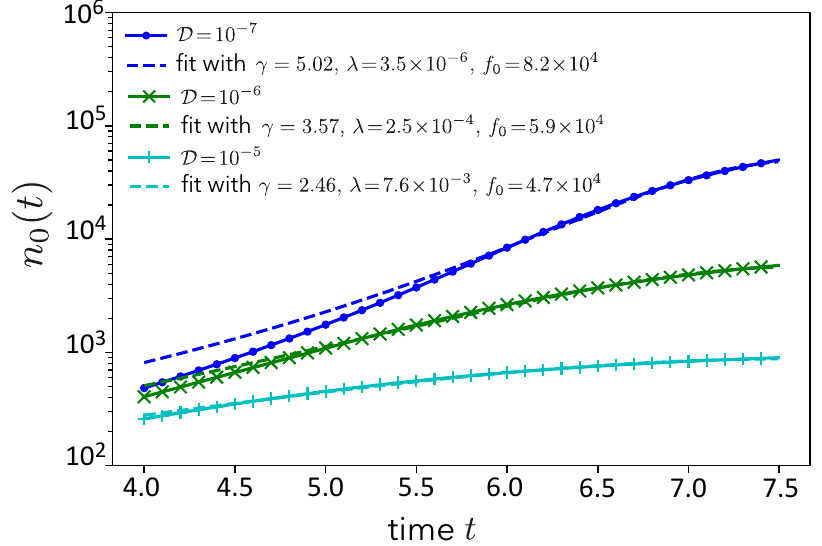}
\caption{
Fits of $n_0(t)$ to Eq.~(\ref{NL_corr}) for three values of $\mathcal{D}$ in the inverse-cascade regime, fixing  $t^* = 6.7$, $\alpha = 2.7$ and $\beta = 2.2$. The obtained fit parameters $\gamma$, $\lambda$, and $f_0$ are given in the legend. 
}
\label{Fig:fit_cascade}
\end{figure}

In the absence of drive ($\mathcal{D} = 0$), the central density $n_0(t)\propto (t^* - t)^{-\alpha}$ diverges around $t = t^*$. In this case, an algebraic fit for $\mathcal{I}=1$ provides $t^* = 6.7$ and $\alpha = 2.7$. A fit of $\epsilon_\text{th}(t)$ to $(t^* - t)^{\beta}$ at intermediate energies, on the other hand, gives $\beta = 2.2$ and therefore $\alpha/\beta = \nu \simeq 1.23$.
For finite $\mathcal{D}$, $n_0(t)$ is non-divergent and is better described by
\begin{equation}\label{NL_corr}
    n_0(t)=\tau(t)^{-\alpha},\ \tau(t)= \big[ f_0(\mathcal{D})^{-1/\alpha} (t^* - t + \lambda(\mathcal{D})t^{\gamma(\mathcal{D})}) \big],
\end{equation}
where we take $\lambda(\mathcal{D})$, $\gamma(\mathcal{D})$ and $f_0(\mathcal{D})$ as fit parameters, fixing $t^* = 6.7$ and $\alpha = 2.7$ for a better accuracy of the fit. 
Once $\tau(t)$ and $\alpha$ are determined, we simply plot $n_\epsilon(t)\tau(t)^\alpha$ as a function of the rescaled energy  $\epsilon\tau(t)^\beta$, with $\beta=2.2$. This leads to the curves in Fig.~3-b of the main text, which exhibit a very good collapse.
Fits of $n_0(t)$ to Eq.~(\ref{NL_corr}) 
are shown in Fig.~\ref{Fig:fit_cascade} for the three values of $\mathcal{D}$ used in the main text. The time window used for the fit is chosen so that an inverse cascade occurs, with too short times  excluded (for $\mathcal{D}=0$, self-similarity typically starts around $t \simeq 5$).

\subsection{Stationary regime}

In the stationary regime, the central density is time independent, $n_0(t) = n_0$, a parameter that nevertheless depends on $\mathcal{D}$. Defining $\smash{\tau(t) = \tau_0 = n_0^{-1/\alpha}}$, the self-similar scaling law reduces to  $n_\epsilon(t)=n_0(\mathcal{D}) f[\epsilon\,n_0(\mathcal{D})^{\beta/\alpha}]$ [which is Eq.~(7) of the main text].
To achieve the collapse of distributions at different $\mathcal{D}$ in Fig.~3-d, we first observe numerically that $n_\epsilon\simeq \epsilon^{-\nu}$ with $\nu\simeq 1.23$ at intermediate energies. Therefore, in this energy range we have $n_\epsilon/n_0(\mathcal{D}) \simeq \epsilon^{-\nu}/n_0(\mathcal{D}) = [\epsilon\,n_0(\mathcal{D})^{1/\nu}]^{-\nu}$. Identification with Eq.~(7) of the main text then gives us $\alpha/\beta=\nu\simeq 1.23$, and a collapse of the curves at different $\mathcal{D}$ when plotting $n_\epsilon(t)/n_0(\mathcal{D})$ as a function of $\smash{\epsilon\, n_0(\mathcal{D})^{1/\nu}=\epsilon/\tau_0^\beta}$.

\subsection{Total energy}

\begin{figure}[h]
\includegraphics[scale=0.6]{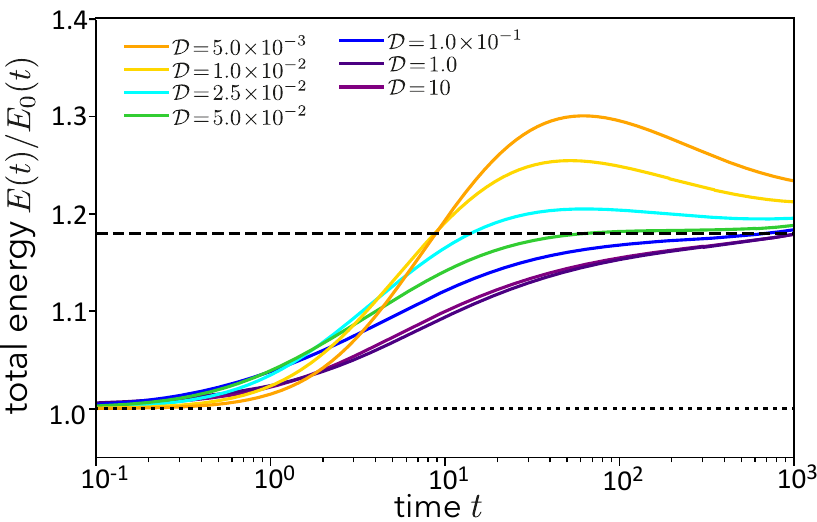}
\caption{\label{Fig:energySM}
Total energy $E(t)=\int_0^\infty d\epsilon\, \epsilon\,\nu_\epsilon n_\epsilon(t)$ 
of the Bose gas as a function of time, normalized to the corresponding energy $E_0(t)$ for $\mathcal{I}=0$. At long times, curves at different $\mathcal{D}$ all converge to a constant $\sim 1.18$.}
\end{figure}

When computing the total energy using the fit function $f_{\text{fit}}(x)$ in the sub-diffusive regime, we obtain $E(t) \simeq 1.19E_0(t)$, $E_0(t)$ being the total energy in the absence of interactions. This result is very close to the exact law $E(t) \simeq 1.18E_0(t)$ [Eq.~(9) of the main text], which is demonstrated in Fig.~\ref{Fig:energySM}.

\section{Experimental data}
\label{Sec:exp}

We finally explain how to extract the coordinates $(\mathcal{D}/\mathcal{I},t\,\mathcal{I})$ corresponding to points in the dynamical phase diagram (Fig.~1 of our main text) from experimental measurements of Ref.~\cite{Martirosyan2024}.

In  Fig.~5-a of Ref.~\cite{Martirosyan2024}, measurements of momentum distributions have been reported in a $^{35}$K gas with $10^5$ atoms confined in a cylindrical trap of radius $R\simeq 15\,\mu$m and length $L\simeq 50\,\mu$m, after an evolution time of $t=1\,\text{s}$, and for a disorder scattering rate $\tau_s^{-1}=15\,\text{s}^{-1}$. A cross-over between a sub-diffusive regime and an inverse turbulent cascade has been identified for an interaction scattering length between $a=20a_0$ and $50a_0$ (with $a_0$ the Bohr radius).
In terms of the dimensionless time $\mathcal{I}\,t=[g^2m^3 \epsilon_0/(2\pi^3\hbar^6)]\times (t\,\epsilon_0/\hbar)$  used in the phase diagram, where $g=4\pi\hbar^2 a/m$ and $\epsilon_0=(\hbar^2/2m)\times(4\pi^2\rho_0)^{2/3}$, these two values respectively correspond to $\mathcal{I}\simeq3.3\times 10^{-5}$ and $2.1\times 10^{-4}$, and to $\mathcal{I}\,t=0.62$ and $3.89$.

Furthermore,  energy measurements carried out for $\tau_s^{-1}=8\,\text{s}^{-1}$ and shown in Fig.~4-a of Ref.~\cite{Martirosyan2024} provide $2E(t)/3\simeq 6.35k_Bt\, \text{nK}$ for the total energy. Comparing with the prediction of our work, $\smash{E(t)\simeq 1.18E_0(t)=1.18\sqrt{\mathcal{D}t}}\times 2\Gamma(5/4)/\Gamma(3/4)$ [see Eq.~(9) of the main text], and using that $\mathcal{D}\propto \tau_s$, we can infer the value of $\mathcal{D}=F^2D/2$ for $\tau_s^{-1}=15\,\text{s}^{-1}$. In terms of the dimensionless units $\hbar/\epsilon_0^3$ used in Fig.~1 of our main text, we find $\mathcal{D}\simeq 4.15\times10^{-8}$. This eventually gives a ratio $\mathcal{D}/\mathcal{I}\simeq1.3\times10^{-3}$ and $2.0\times 10^{-4}$ for $a=20a_0$ and $50a_0$, respectively.


\begin{thebibliography}{99}
\bibitem{footnote}
In the three regimes of inverse cascade, metastable equilibrium and sub-diffusion, the prefactor in $\tau(t)$ is chosen so that the scaling function $f(0)=1$.
\bibitem{footnote2}
The factor 1.18 is recovered if for $f(x)$ one uses the exact fit function given in the legend of Fig.~\ref{Fig:collapse} instead of Eq.~(\ref{eq:subdiffexp}).
\end{thebibliography}
\end{document}